\documentclass{aa}
\usepackage{graphics,supertabular}
\usepackage{longtable}
\usepackage{graphicx}
\usepackage{amsfonts}
\usepackage{natbib}
\usepackage{amssymb}

\begin{document}

\title{A search for OB stars in the field of the galactic OB association Bochum 7}
\subtitle{II. Proper motion and IR photometry\footnote{Table n$^{\circ}$ 3 is only available in electronic form
at the CDS via anonymous ftp to cdsarc.u-strasbg.fr (130.79.128.5)
or via http://cdsweb.u-strasbg.fr/cgi-bin/qcat?J/A+A/}}
\author{M. A. Corti\inst{1,2} 
\and R. B. Orellana\inst{2,3} \and G. L. Bosch \inst{2,3}}
\authorrunning {Corti, Orellana \& Bosch}
\titlerunning{Bochum 7 - Proper Motion and IR Photometry}

\institute{Instituto Argentino de Radioastronom\'{\i}a (CCT-La Plata, CONICET; CICPBA),
 C.C. No. 5, 1894 Villa Elisa, Argentina
\and 
Facultad de Ciencias Astron\'omicas y Geof\'{\i}sicas,
Universidad Nacional de La Plata, Paseo del Bosque s/n,
1900 La Plata, Argentina
\and 
Instituto de Astrof\'{\i}sica de La Plata (CCT-La Plata, CONICET), Argentina}

\date{Received \today; accepted \ldots }
\abstract
{}
{We plan to identify the members of the Bochum\,7 association by performing simultaneous astrometric and spectrophotometric analyses, and estimate its distance and evolutionary stage.}
{We used our own visual spectroscopic and UBV photometric data of a 30' $\times$ 30' region centered at $\alpha$ = 8$^h$44$^m$47.2$^s$, $\delta$ = - 45$^{\circ}$58\arcmin55.5\arcsec . This information enabled us to estimate the spectral classification and distance of all stars present in the region. The proper motion was analyzed with data of the {\rm UCAC\,5} catalog and was used to identify the members of this association. We added {\rm JHK} data from {\rm 2MASS} and {\rm IRAS} catalogs to check for the presence of infrared (IR) excess stars}
 {We found that Bochum\,7 is an OB association with at least 27 identified stellar members ($l = 265^{\circ}\!\!$.12, b = $-2^{\circ}\!\!$) at a distance of $\simeq$ 5640 pc. Its proper motion is $\mu_{\alpha}cos\delta$ = $-4.92 \pm 0.08$ mas yr$^{-1}$,  $\mu_{\delta}$ = $3.26 \pm 0.08$ mas yr$^{-1}$. We derived an average heliocentric radial velocity of $\sim$ 35 km s$^{-1}$ and were able to confirm the binary nature of the (ALS\,1135) system and detect four new binary star candidates. Analysis of data for massive Bo\,7 star candidates points towards a young age ($ \leq 3\,\times$ 10$^6$ years old) for the association, although the presence of a previous episode of star formation remains to be analyzed.
 }
{} 


\maketitle
%

\section{Introduction}

The need for relevant statistics, such as the determination of
large-scale galactic structure and the quantification of metallicity gradients on the disk 
require the knowledge of OB associations at distances of several kiloparsecs.

The study of associations located at distances greater than the kiloparsec becomes extremely complex
and the use of a single research technique yields unreliable results with large uncertainties. Bochum7 (Bo 7) is located in the  region known as the "Puppis Window" in the third quadrant of our Galaxy. The low interstellar extinction in this region allows for reliable observational data to be obtained for stars that belong to the Perseus arm.

The available information can be summarized as follows:
\begin{itemize}
\item[-] \citet{ste71} published the Luminous Stars in the Southern Milky Way (ALS) catalog.
\item[-] \citet{mof75} performed a survey of open clusters with photoelectric photometry using the 61\,cm telescope of the Bochum University at La Silla, Chile. They found that nine members of the ALS catalog possibly formed an association with $\overline{E(B-V)}$ = 0.86 and $\overline{d}$ = 5.8 kpc.
\item[-] \citet{lun84} evaluated existing data about Wolf-Rayet (WR) stars in OB associations and open clusters. They confirmed the membership of WR12 to Bochum\,7, as originally proposed by \citet {nie82} and estimated a 30 pc diameter for the association.
\item[-] \citet{sun99} used UBV and H$\alpha$ photometry and located Bo\,7 at a distance of 4.8 kpc near the center of the Vela\,OB1 stellar association \citep{ree00}. They estimated the age of the association to be about 6 Myr old.
\item[-] \citet{dia02} published the new catalog of optically visible open clusters and candidates. For this, they worked with the Tycho\,2 data and presented 953 members of Bo\,7 with $\mu_{\alpha}cos\delta$ = -0.05 mas yr$^{-1}$ and $\mu_{\delta}$ = 3.34 mas yr$^{-1}$, although no detailed breakdown of the method was presented. 
\item[-] \citet{arn07} based on the analysis of the neutral hydrogen (HI) supershells carried out at the Southern Galactic Plane Survey (SGPS) proposed that the OB-association Bo\,7 was born as a consequence of the evolution of GS263-02+45. 
\item[-] \citet{cor07} performed the first spectroscopic analysis with the OB star candidates selected from $UBV$ aperture photometry of the 30 arc-minute field centered at $\alpha$ = 8$^h$44$^m$47.2$^s$, $\delta$ = - 45$^{\circ}$58\arcmin55.5\arcsec (J2000.0) (hereafter Paper\,I).
\item[-] \citet{mic13} performed a photometric investigation focused on the binary system ALS\,1135 and discovered 17 variable stars in the field of Vela\,OB1 or Bo\,7 associations.
\end{itemize}

Stellar proper motions constitute a powerful tool for the identification of members of stellar groupings, open clusters, and OB associations \citep[and references therein]{dia02, deze99}. It is therefore necessary to have stellar catalogs containing very good proper motions and covering a large extent of the celestial sphere down to faint magnitudes. The goal of this work is to increase our knowledge of the stellar components of Bo\,7. For this, we now present an astrometric study that makes use of the UCAC\,5 \citep{zac17} catalog. With this survey we were able to derive the proper motion of stars located in the field of the Bochum\,7 association.
The combination of the astrometric results with the analysis shown in Paper I yield the astro-spectrophotometric members of the Bo 7 association, its distance, age, radial velocities, and so on. 
In this way, it is possible to increase our knowledge of the galactic region where the Bo 7 association is located.

The paper is organized as follows: Section 2 describes the available data and the analysis methods,  results are shown in Sect. 3, and the discussion about global properties of the association and individual stars is developed in Sect. 4. Section 5 includes a brief summary of our findings. Additional material such as detailed information from radial velocity measurements for individual stars is included in an Appendix.

\section{Observational data}
\subsection{Astrometric data}

The astrometric data used in this paper, stellar position and proper motion, are extracted from the UCAC\,5 catalog.

The UCAC\,5 catalog presents the positions of over 180 million stars, and  proper motions for 107.7 million of them. The  UCAC\,5  positions are on the Gaia DR1 \citep{brow16,pru16}
reference system. The positions have mean epochs around 1998 to 2000 in the south, and around 2001 to 2003 in the north. Formal position errors are about 8 mas for stars with magnitude 11 and 20 mas at magnitude 14. The UCAC\,5 catalog presents very precise proper motions combining UCAC\,5 with Gaia DR1. Errors in proper motions of bright stars ($R$ = 11 to 15 mag) lie between 1 and 2 
mas yr$^{-1}$ and can increase to  5 mas yr$^{-1}$ for the fainter stars.

A detailed description about the construction of the UCAC\,5 catalog can be found in \citet{zac17}. Data extraction was performed using the SIMBAD Astronomical Database (CDS).

\subsection{Photometric and spectroscopic data}
\label{photspect}
The $UBV$ photometric data analyzed by us in order to investigate the OB association Bo\,7
are those presented in Sects. 2 and 3.1 of Paper\,I. We selected stars with $V <$ 15 mag and reddening$-$free parameter $Q < -0.3$ \citep{lando82} in the 30' $\times$ 30' photometric image centered at $\simeq \alpha$ = 8$^h$45$^m$, $\delta$ = $- 45^{\circ}$59' ($l = 265^{\circ}\!\!$.12, b = $-2^{\circ}\!\!$.0) and found 248 probable members of this association (see Table 3). The photometric study was complemented with $JHK$ data from the Two Micron All Sky Survey (2MASS) \citep{skr03}

We also investigate the IRAS point sources present in the region under study consulting the Gator Catalog Query\footnote{http://irsa.ipac.caltech.\-edu\-/missions} in the IRAS Point Source Catalog v2.1 (PSC) section. 

Spectral images data studied in this work are those presented in Sect. 3.2.1 of Paper\,I. 

\begin{figure}
\centering
\includegraphics[width =0.5\textwidth]{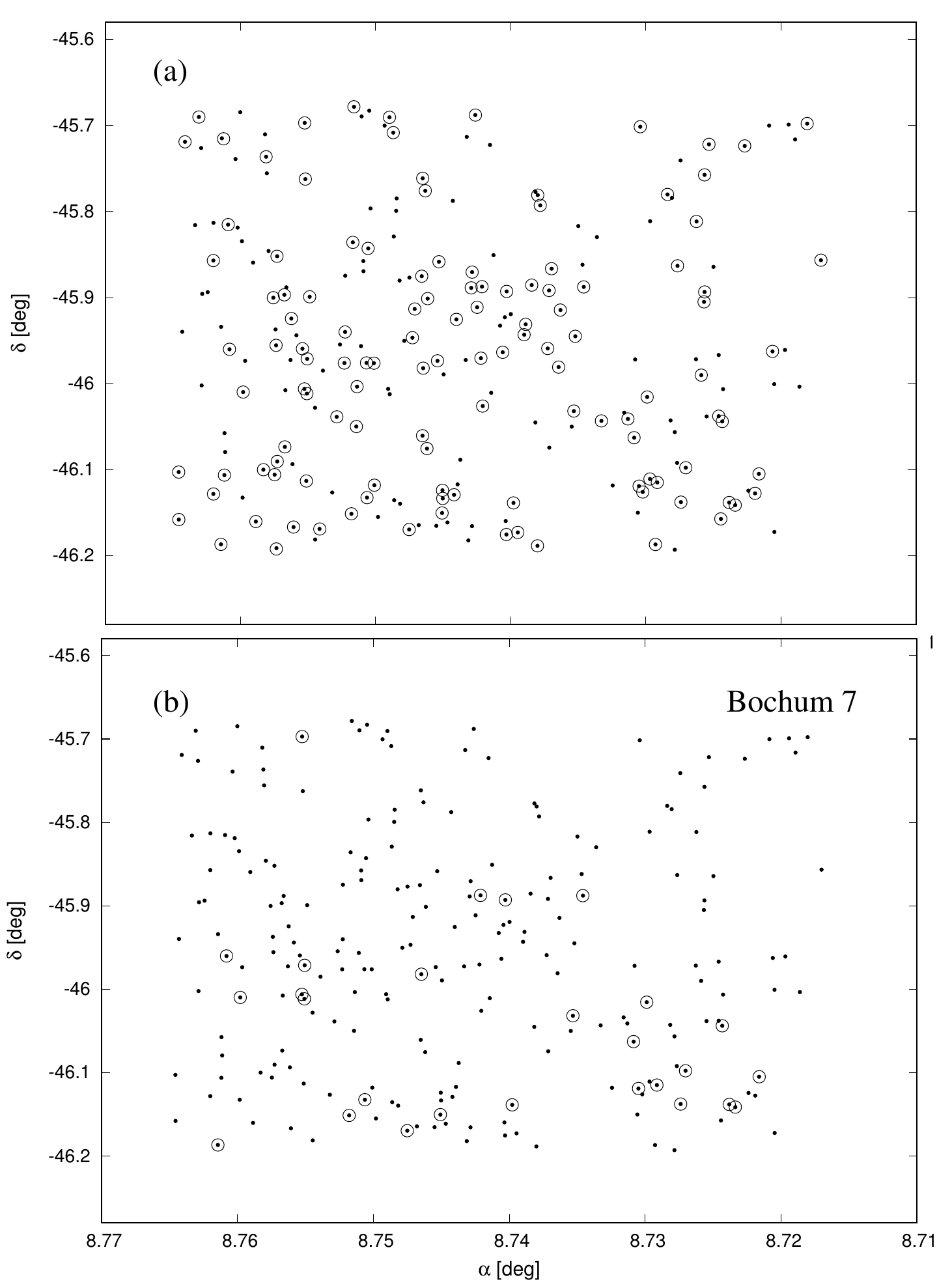}
\caption{Spatial distribution of the stellar association. a) Open
circles around the dots represent the 123 stars with similar proper motions identified in the first analysis; the points represent the rest of the stars of the region; a total of 219 stars. b) {Open circles represent the 27 astrometric and spectrophotometric Bo 7 association members.} All of them were derived using the model proposed by \citet{ore10}. }
\label{fig:Coord2CatVela}
\end{figure}

\section{Results}
\subsection{First astrometric analysis}
\label{astrom}

Although impossible to detect by visual inspection of direct imaging, the existence of an association can be inferred from their proper motions, through the analysis of the Vector Point Diagram (VPD). In this diagram the relatively small differences of the velocities of individual stars from the association can be detected as an over-density. As a first step, we identified in UCAC\,5 catalog the 219 stars out of the 248 candidates proposed as members of Bo7  (Table 3). For its identification, the astrometric catalog UCAC4 was used as an intermediate catalog in the following way. From the 248 early-type stars present in the field of view of Bo 7, 226 were identified in UCAC4 from position ($\alpha$, $\delta$) and magnitudes (V, B). From the UCAC4 sample, 219 were further identified in the UCAC5 from position ($\alpha$, $\delta$) and 2MASS magnitudes (J, H, K.). Subsequently we analyzed the presence of an over-density with the model proposed by \citet{ore10}. In their paper, they suggested for the first time, a standardized way to identify the members of an association. To do this, the authors applied the same technique used for open clusters considering only the proper motions based upon the maximum likelihood principle \citep{san71}.

The distribution of the proper motions will be the overlapping area of two bivariate normal frequency functions in an elliptical subregion of the VPD selected as indicated by \citet{vas65};

\begin{eqnarray}
\label{phit}
\Phi_i(\mu_{xi},\mu_{yi}) = \phi_{1i}(\mu_{xi},\mu_{yi}) + \phi_{2i}(\mu_{xi},\mu_{yi})
,\end{eqnarray}

where $\phi_{1i}$ is a circular distribution for cluster stars and $\phi_{2i}$ is an elliptical  distribution  for  field  stars.
The circular and elliptical distributions take the following form.

\begin{eqnarray}
\label{phi2}
&&\phi_{1i}(\mu_{xi},\mu_{yi})= \frac{N_a}{ 2\pi\sigma_{a}^2}\times \nonumber\\
&&\times\exp\left[-\frac{(\mu_{xi}-\mu_{xa})^2 + 
(\mu_{yi}- \mu_{ya})^2}{2\sigma_{a}^2}\right]
,\end{eqnarray}

and
\begin{eqnarray}
\label{phi3}
&&\phi_{2i}(\mu_{xi},\mu_{yi})= \frac{N_f}{ 2\pi\sigma_{xf}\sigma_{yf}}\times\nonumber\\
&&\times\exp\left[-\frac{(\mu_{xi}-\mu_{xf})^2}{2(\sigma_{xf})^2} - 
\frac{(\mu_{yi}- \mu_{yf})^2}{2(\sigma_{yf})^2}\right]
,\end{eqnarray}

where the symbols $\sigma_{xf}$, $\sigma_{yf}$ are the elliptical dispersions for the field stars, $\sigma_a$ the circular dispersion for the association stars, $\mu_{xf}$, $\mu_{yf}$ the field star mean proper motion, and 
$\mu_{xa}$, $\mu_{ya}$ the association mean proper motion. $N_a$ is the number of association members, and $N_f$ the number of field stars. These parameters are found by applying the maximum likelihood principle. Once determined, the  probability for the i$-$th star was calculated as
\begin{equation}
\label{Pc}
P_i(\mu_{xi},\mu_{yi})= \frac{\phi_{1i}(\mu_{xi},\mu_{yi})}{\Phi_i(\mu_{xi},\mu_{yi})}.
\end{equation}
A cluster member is found when $P_i \geq 0.5$.

In this procedure we do not consider the influence of proper motion errors. \citet{che97} and \citet{ore14} show that the error does not change the parameters significantly and more than 80\% of the members maintain their condition.

The astrometric study begins by applying the maximum likelihood method to an elliptical subregion of the VPD containing 186 stars (Fig.~\ref{fig:VPDUCAC5}a). 
The parameters of the stars with similar proper motions obtained from this method are
 $N_a$ = 123,  $\mu_{xa}$ = $- 5.42 \pm 0.06$ mas yr$^{-1}$,  $\mu_{ya}$ = $3.63 \pm 0.06$ mas yr$^{-1}$,  $\sigma_a$ = $0.74 \pm 0.05$ mas yr$^{-1}$. Figures \ref{fig:Coord2CatVela}(a) and \ref{fig:VPDUCAC5}(b) show the location of the 123 stars in the spatial distribution and in the VPD, respectively, with open circles around the dots. In order to identify the member stars of Bo 7, we analyzed the distance modulus (DM) of the 123 possible members.
 
\subsection{Spectral types and distances}
\label{sptanddist}

The spectral classification of stars in this study were obtained in Paper\,I and are listed in Table \ref{PhotBo7}.

In this paper, the DM of each star was obtained using Eq. (4) of Paper\,I. The intrinsic $(B-V)_0$ color index and $A_V$ visual extinction were calculated from our observational data with Eqs. \ref{intrinsicol} and \ref{absorption}, respectively. The latter were obtained from \citet{lando82}, together with the absolute visual magnitude of each star according to its spectral type. 
\begin{equation}
\label{intrinsicol}
(B-V)_0 = 0.319 \times (U-B) - 0.23 \times (B-V) - 0.026
,\end{equation} 
\begin{equation}
\label{absorption}
A_V = 3.23 \times [(B-V) - (B-V)_0]
.\end{equation}

For stars with two different spectral types, we calculated the average value of $M_v$ and interpolated linearly whenever needed. The error in the determination of the distance to each star was found to be around 30\%. We obtained that value with errors propagation. For this, we considered average magnitude errors of 0.08 in $V$, 0.10 in $B$, and 0.18 in $U$ (Paper\,I) and an uncertainty of 0.5 in the estimation of $M_v$ \citep{wal72}. 
Although errors in individual distances derived from spectroscopic classification are large, they can still provide a first-order estimation useful for removing foreground and background stars. We finally chose to consider possible members of the  Bo\,7 association to those stars whose distance matched with the average distance ($\overline{\mathrm{DM}}$ = 13.725) obtained from the subset of ALS stars listed as Bo 7's members by \citet{mof75}, accounting for an uncertainty of 0.5 mag. 
In this way, were considered 26 stars with 13.475 $\leq$ DM $\leq$ 13.975 resulting from the first astrometric analysis as probable members of Bo 7. We also included the binary system ALS\,1135 even though its photometric distance (DM = 14.1) is slightly off the range as binary pairs have larger uncertainties in their photometric distance determinations.
The results obtained in the calculation of DM are shown in Table \ref{PhotBo7}. The DM obtained for stars in common with Paper\,I are in very good agreement.

\subsection{Final astrometric analysis and association distances} 
\label{ucac5catal}

In order to improve the identification of the members of Bo 7, the maximum likelihood method was applied to a group of stars selected from the 186 stars of the elliptical subregion and DM analysis as follows.

\begin{figure*}
\centering
\includegraphics[width =0.6\textwidth]{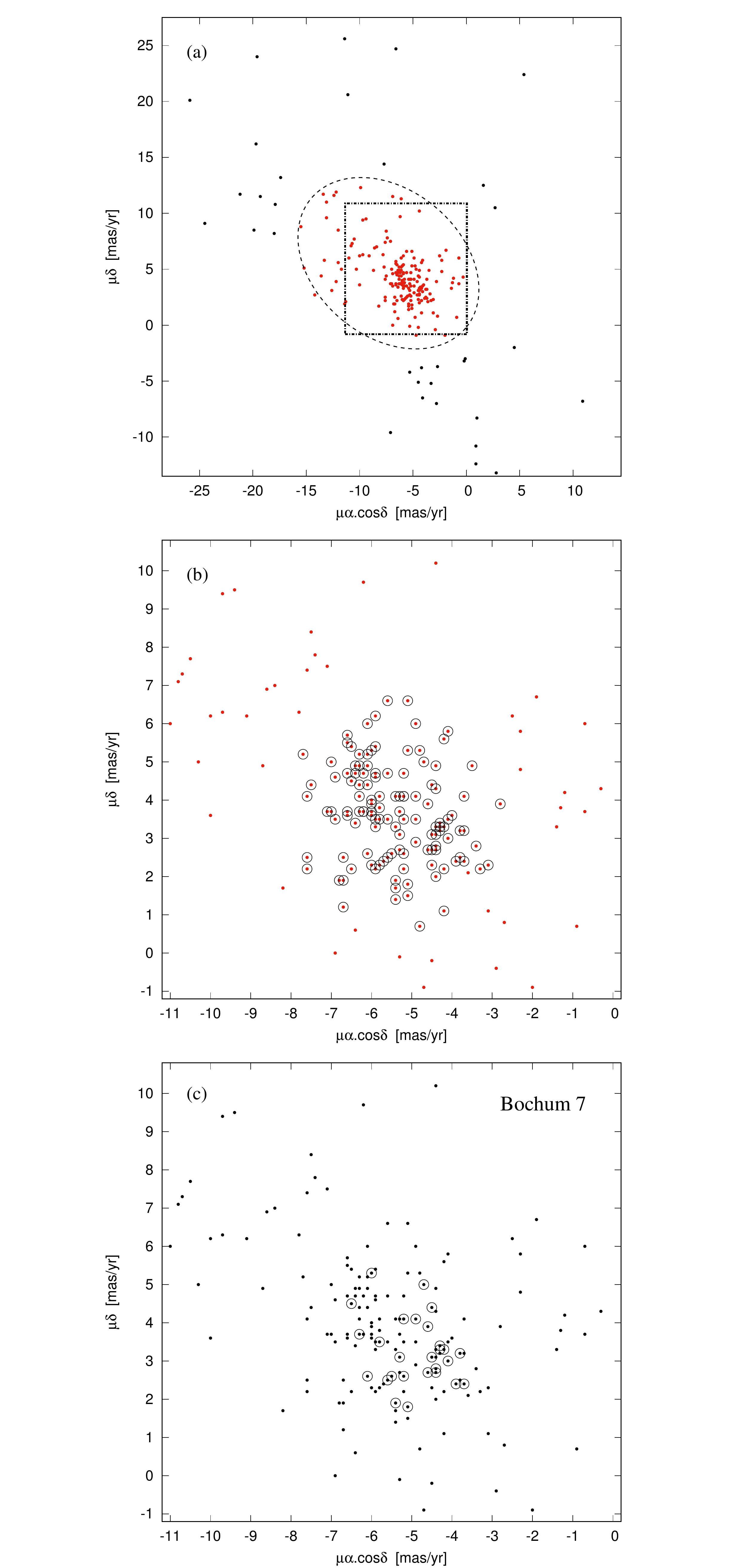}
\caption{VPD of the stellar association consulting the UCAC5 catalog. (a) Maximum likelihood method assigns 
the Gaussian distribution bivariate elliptic to the 186 field stars. Points outside ellipse represent the rest of the region's stars
in the first analysis. (b) A zoom image of the square indicated in (a). Open circles around the dots represent the 123 astrometric members that the maximum likelihood method assigns 
the Gaussian circular distribution bivariate. (c) Open circles represent the 27 astrometric and spectrophotometric association Bo 7 members. All of them were obtained using the model proposed by \citet{ore10}.}
\label{fig:VPDUCAC5}
\end{figure*}

We removed the contamination produced by 94 stars located in the foreground and background of the Bo 7 sample, that is, those with DM in the ranges 13.475 $>$ DM $>$ 13.975 obtained following the procedure outlined in Sect. \ref{sptanddist}. The \citet{ore10} model was applied again to these 92 stars (186 (stars of the elliptical subregion) -  94 (stars outside the Bo7 bin)) and the refined parameters found were $N_a$ = 27,  $\mu_{xa}$ = $-4.92 \pm 0.08$ mas yr$^{-1}$,  $\mu_{ya}$ = $3.26 \pm 0.08$ mas yr$^{-1}$,  $\sigma_a$ = $0.42 \pm 0.07$ mas yr$^{-1}$. In this case, 27 stars 
were identified as likely candidates to be members of Bo 7. Figures~\ref{fig:Coord2CatVela}(b) and ~\ref{fig:VPDUCAC5}(c) show their spatial distribution and VPD, respectively. Table \ref{bo7astrom} lists the stars found with this procedure. 

After combining membership by proper motions and distances via spectrophotometry, the distance moduli of the Bo 7 association is found to be 13.7 $\pm$ 0.2 (5.6 $\pm$ 1.7 kpc).

\begin{table*}[ht!]
\begin{center}
\leavevmode
\caption{Possible astrophotometric members of Bo 7.}
\label{bo7astrom}
\begin{scriptsize}
\begin{tabular}{cccccc}
\hline
\hline

ID$^{(1)}$ & $\alpha_{J2000.0}$ &   $\delta_{J2000.0}$   & UCAC\,5 & $\mu_{\alpha}cos{\delta}$  &  $\mu_{\delta}$ \\
  & $(h:m:s)$ & (${\circ}$:':'')   &   &   (mas yr$^{-1}$) &  (mas yr$^{-1}$) \\
\hline

343 & 8:45:41.2 & -46:11:13  &  532984646920682 & -5.8 & 3.5  \\
390 & 8:45:38.9 & -45:57:36 &  532994570012623 & -6.0 & 5.3 \\
521 & 8:45:35.3 & -46:00:35 &  532994535653243  & -5.5 & 2.6  \\
1031 & 8:45:18.9 & -45:41:49 & 532996518209937 & -6.3 & 3.7  \\
1052 & 8:45:18.9 & -46:00:22 & 532986340855840 & -4.6 & 3.9 \\
1069 & 8:45:18.2 & -45:5817 & 532986358035711 & -5.2 & 4.1 \\
1071 & 8:45:18.3 & -46:00:42 & 532986340855840 & -6.5 & 4.5 \\
1470  & 8:45:06.4 & -46:09:05 &  532984921798650 & -4.5 & 4.4  \\
1607 & 8:45:02.2 & -46:07:57 & 532984928670556   &  -5.4   & 1.9   \\
1939 & 8:44:51.0 & -46:10:11 & 532984873695016  &  -4.6   & 2.7  \\
2041 & 8:44:47.3 & -45:58:55 & 532986423319214 & -4.4 & 2.8 \\
2197 & 8:44:42.3 & -46:09:01 & 532986065977889 & -5.6 & 2.5 \\
2511 & 8:44:31.6 & -45:53:15 & 532986842508021  & -5.3 &  3.1  \\
2720 & 8:44:25.1 & -45:53:35 & 532986859687891 & -5.1& 1.8  \\
2781 & 8:44:23.2 & -46:08:19 & 532985584941600   & -4.3 & 3.3 \\
3211 & 8:44:07.1 & -46:01:55 & 532986536706341 & -5.2 & 2.6 \\
3273 & 8:44:04.5 & -45:53:16 &  552202093686894 & -4.4 & 2.7  \\
3669 & 8:43:51.1 & -46:03:46 & 532986014438328 & -3.7 & 2.4 \\
3705 & 8:43:49.8 & -46:07:09 & 532985883871323 & -4.2  &        3.3 \\
3767 & 8:43:47.6 & -46:00:56 & 532986045362086 & -4.7 & 5.0 \\
3829 & 8:43:44.9 & -46:06:54 & 532985880435349 & -4.5 & 3.1 \\
3982 & 8:43:38.6 & -46:08:16 & 532985729252501 & -4.3 &  3.4 \\
4009 & 8:43:37.3 & -46:05:52 & 532985959462720 & -3.9 & 2.4 \\
4234 & 8:43:27.6 & -46:02:38 & 552201354523083 & -4.1  & 3.0  \\
4288 & 8:43:25.7 & -46:08:17 & 532985818587821 & -4.9 & 4.1 \\
4331 & 8:43:24.1 & -46:08:29 & 532985815581938 & -3.8 & 3.2 \\
4476 & 8:43:17.8 & -46:06:18 & 552201182724391 & -6.1 & 2.6 \\
\hline
 \end{tabular}

 \parbox{14cm}{{\scriptsize $^{(1)}$ This paper}}\\
\end{scriptsize}
\end{center}
\end{table*}

\subsection{Stellar radial velocities}
\label{stellarrv}

The radial velocity is another important parameter to be studied for each star member of the association. We gathered radial velocity determinations for 15 stars in the Bo 7 association (Column 9 of Table \ref{PhotBo7}). A detailed breakdown of individual measurements can be found in Appendix A.

A single determination of radial velocities of massive stars must be handled with care and the uncertainties in individual distances are large. Similarly, we perform an individual check on their velocities and those expected according to the Milky Way rotation. 

\begin{figure}[!ht]
\centering
\includegraphics[width=0.5\textwidth]{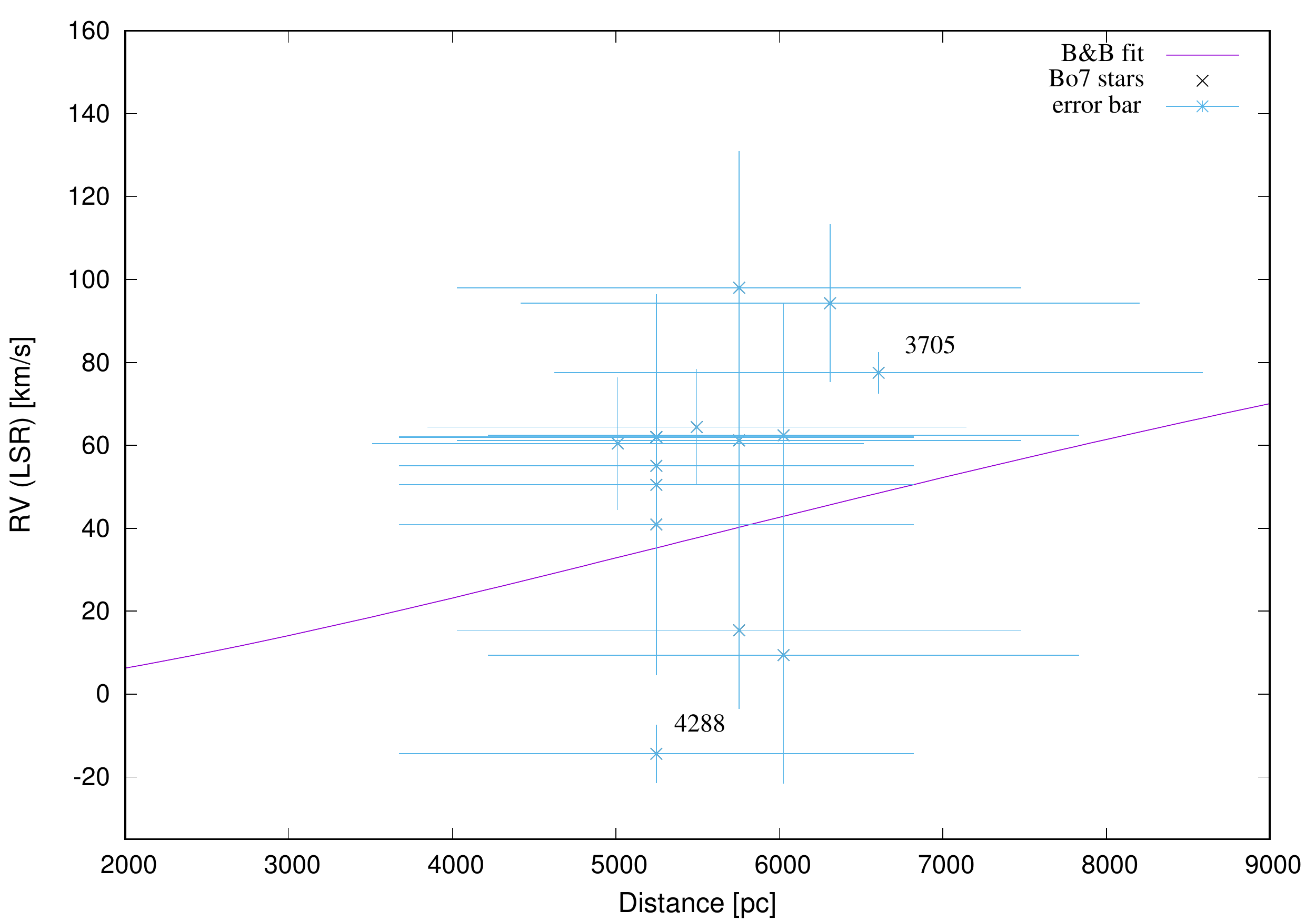}
\caption{LSR corrected radial velocity vs. distance of Bo\,7 member candidates plotted as blue crosses with their corresponding error bars. The pink curve displays the fit of the galactic rotation model applied to the third quadrant of our Galaxy \citep{bra93}. Star \#4288 is the only member more than 3$\sigma$ away from its expected location (discussed further below). As a reference,  star \#3705 is the ALS\,1135 binary system.}
\label{RotGalaxy}
\end{figure}

Figure \ref{RotGalaxy} shows the LSR radial velocity, RV$_{LSR}$, and the distance from each star to the Sun; both parameters are included with their respective error bars. The fit of the Galactic rotation curve model by \citet{bra93} applied to the third quadrant of the Galaxy (see \citet{san17}) is also plotted as a reference, although the \textit{uncertainties} in the model determination at about 10 kpc from the Galactic Centre are large. Large uncertainties involved in the individual parameters are reflected in the observed scatter, and only star \#4288 is located farther than expected for its errors. Appendix A includes two radial velocity determinations with similar low values, making it the strongest candidate to be checked for RV anomalies in the future with higher-resolution spectra.

The galactic rotation model (Fig. \ref{RotGalaxy}) indicates that at a distance to the Sun of $\sim$ 5.7 kpc ($\sim$ 13.7 mag) the Local Standard of Rest radial velocity (V$_{LSR}$) value is $\sim$ 40 km s$^{-1}$ (V$_{he}$ $\simeq$ 35 km s$^{-1}$). This is in good agreement with the radial velocity values with their errors shown for Bo 7 members in Table \ref{PhotBo7}. 

\section{Discussion}
\label{the_associations}

\subsection{Recent membership identifications.}

\citet{mic13} performed a photometric and spectroscopic investigation focused on the ALS\,1135 binary system and presented a photometric analysis of the surrounding region. They built a $VI$ color magnitude diagram and used additional information from spectroscopic classification found in the literature and variable stars identified from their photometry to identify six stars in this region as possible Bo\,7 members. Five of these stars are included in the present work and we agree on member candidacy for stars \#3705 (ALS\,1135), \#3669 (ALS\,1137) and \#3829 (CBN\,84344.7-460656), although we reject CBN\,84348.6-460736 and CBN\,84346.7-460641, which, according to our astrometric and spectrophotometric study, would not be identified as Bo\,7 members in terms of their distances.

\subsection{Size and evolutionary age.}

Defining the boundary of the association is a nontrivial issue. The lack of a visible concentration of stars among the background forces us to distinguish this group kinematically, but although proper motion surveys are available, it is not possible to derive radial velocities for all the stars. Furthermore, chances of massive stars being part of a multiple system are high \citep{bos09}, but it is difficult to confirm whether or not a star belongs to a multiple system when there are only a few spectra available for it. 

As was explained in Sect.~\ref{astrom}, with data of proper motion it is possible to make a first selection of association members, however with these data alone we cannot define the size of them. With a distance of $\simeq$ 5640 pc adopted for the Bo7 association, our field of view limits the surveyed area to a 50 pc-wide region. This size corresponds to the typical sizes of the stellar associations \citep{efr87} although it is somewhat larger than the diameter of 30 pc found by \citet{lun84}. This association is projected onto the outer border of the HI supershell $GS263-02+45$ \citep{arn07}. Both objects have similar distances and radial velocities, so it is possible that a physical link between them is present.

\begin{figure}[!t]
\centering
\includegraphics[width=0.5\textwidth]{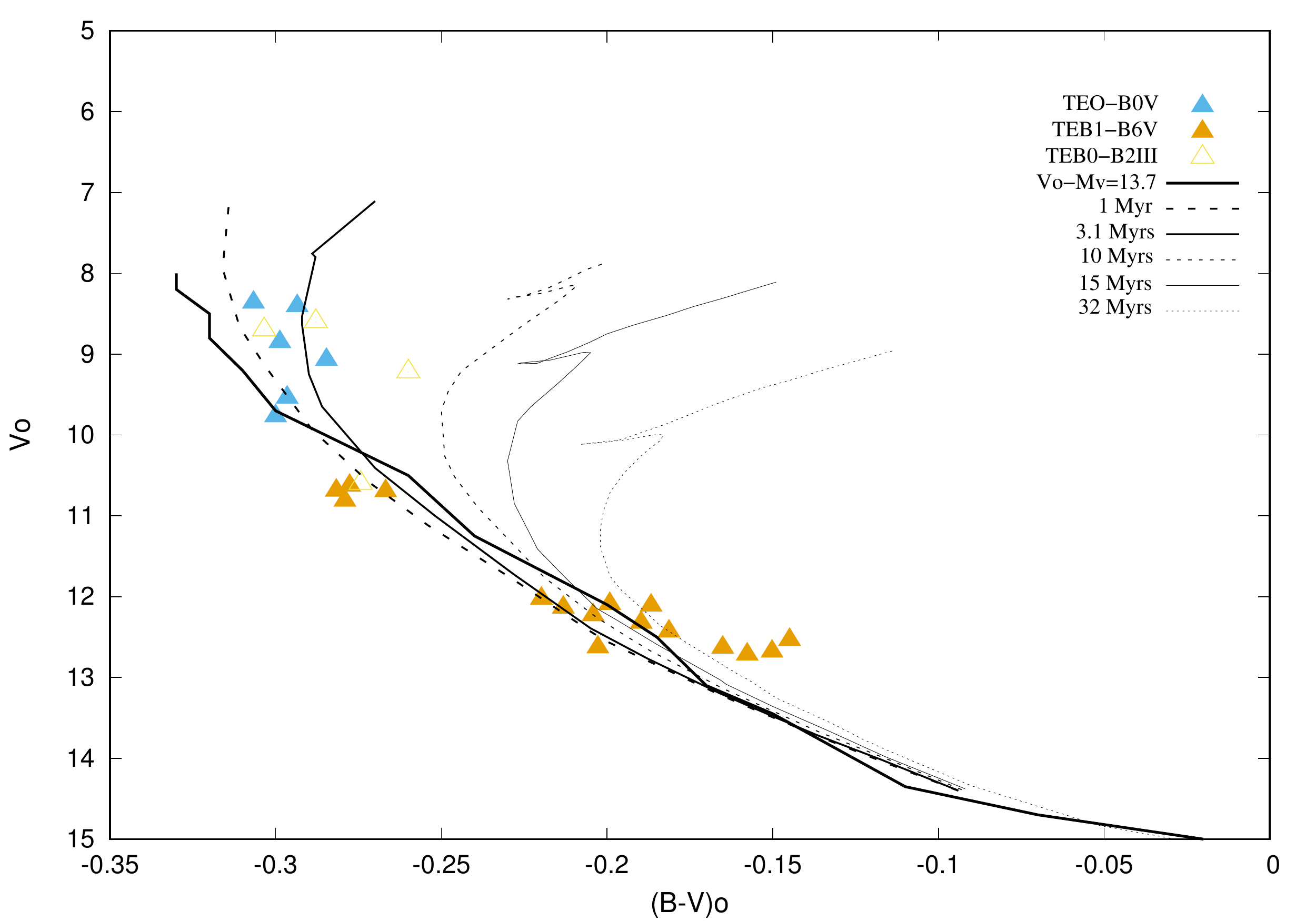}
\caption{CMD of member stars of Bo7. The MS stars among O6.5 and B0 are identified with light blue triangles, the MS stars among B1 and B6 are identified with yellow triangles, and the luminosity class-III stars among B0 and B2 are identified with yellow open triangles. The thicker continuous curve is the \citet{lando82} HR diagram for early stars; the other curves are \citet{mari08} isochrones for $z$ = 0.02. All the reference curves are corrected by an apparent distance modulus of 13.7.}
\label{cmbo7}
\end{figure}

By combining the photometric and spectroscopic information on member candidates, we can decipher the evolutionary stages of the association. 
Figure~\ref{cmbo7} presents the color-magnitude diagram (CMD) of Bo 7. This figure shows the intrinsic visual magnitude and the intrinsic color index of each star and stars are labeled according to their luminosity class. The theoretical isochrones for solar metallicity, mass loss, and overshooting \citep{mari08} are also plotted as a reference. 
Figure~\ref{cmbo7} shows that main sequence (MS) and giant stars occupy a particular region of the CMD, with an unavoidable scatter, possibly due to stars from different formation events that belong to multiple systems, stars with emission lines, or fast rotators. As we are dealing with an incomplete sample of the association, we do not have enough stars to attempt a precise fitting of isochrones in order to discriminate different populations among them. However, it is evident that almost all massive stars lie close to the 3 $\times$ 10$^6$ year-old isochrone, pointing towards very recent episode(s) of star formation, younger than 1 $\times$ 10$^7$ years old. The presence of WR12, a Wolf-Rayet star in the field, can also be linked to recent massive star formation and was the key driver of the original spectrophotometric study.   The spectral classification of WR12 (WN8h), which has been
identified as a single spectrum spectroscopic binary, suggests an evolutionary age of 3.5$\times 10^6$ years \citep{smi08}. \cite{vandh01} estimates a distance of 5 kpc based on photometric data and an absolute magnitude $M_v$ of -5.48. However, we must always keep in mind that magnitudes of WR stars are very uncertain. The very well studied star GR290 \citep{pol16} presents a visual magnitude that ranges from 18.3 to 18.7 in its WN8h phase. Regarding its proper motion, WR12 has a $\mu_{\alpha}cos{\delta}$ of -4.4 mas yr$^{-1}$ and $\mu_{\delta}$ of 2.8 mas yr$^{-1}$, coincident with the proper motion values of the Bo7 members. 

\begin{figure}[!t]
\centering
\includegraphics[width=0.5\textwidth]{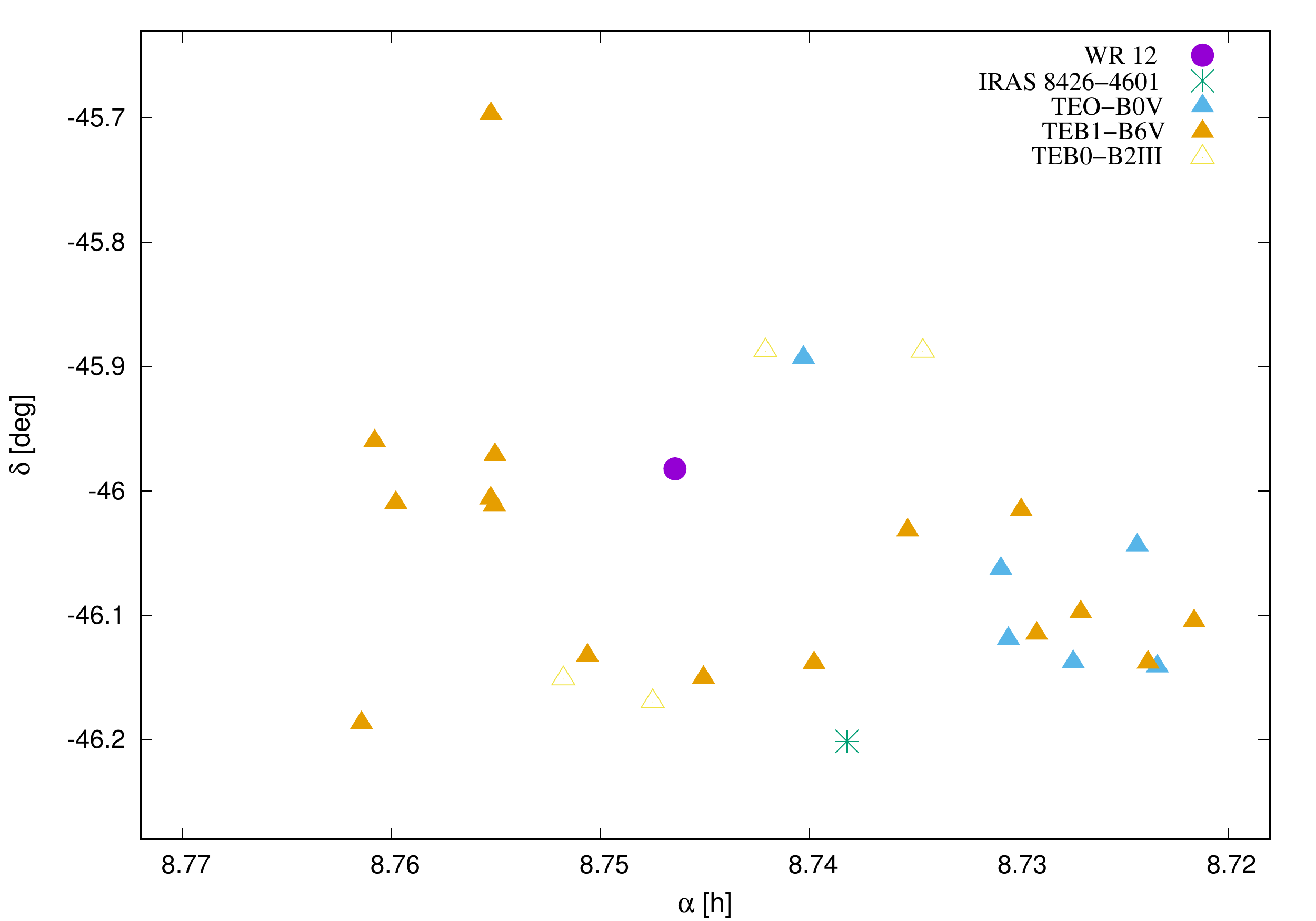}
\caption{Spatial distribution of the Bo 7 members. The WR12 star is identified with a large lilac dot and the IRAS source is identified with a green asterisk. The other symbols are the same as those in Fig. \ref{cmbo7}.}.
\label{Bo7TE}
\end{figure}

Figure \ref{Bo7TE} shows the spatial distribution of the Bo7 members, split into three groups according to their spectral types. One group includes the MS stars between O\,6.5 and B\,0 type (light blue triangles) which can be linked to the most recent and massive formation episode. The other group includes the MS stars from B\,1 to B\,6 (yellow triangles), and the last group includes the giant stars from B\,0 to B\,2 (yellow open triangles) encompassing evolved and less massive stars.
The young and massive group seems more concentrated towards the southwestern region of the surveyed area. Furthermore, in the southern edge, we can identify the infrared (IR) source IRAS\,08426-4601 mentioned in Sect. \ref{photspect}, which can be associated with ongoing star formation in that area. The relatively older and less massive population is more evenly spread throughout the field covered by our photometry. This scenario, where different evolutionary stages seem to occupy different areas, can be interpreted as a trace of sequential star formation, but as stated by \cite{far09} when discussing the N159/160 region of the LMC, a detailed and more complete study is needed before any robust conclusions can be made. Furthermore, it is worth recalling the possible scenario described by \cite{2011sca..conf...85B} where stars could be formed in loose filamentary groups and then disperse in the background without the need of a large cluster with subsequent gas expulsion.

\subsection{Infrared photometry}

\begin{figure}[!ht]
\centering
\includegraphics[width=0.5\textwidth]{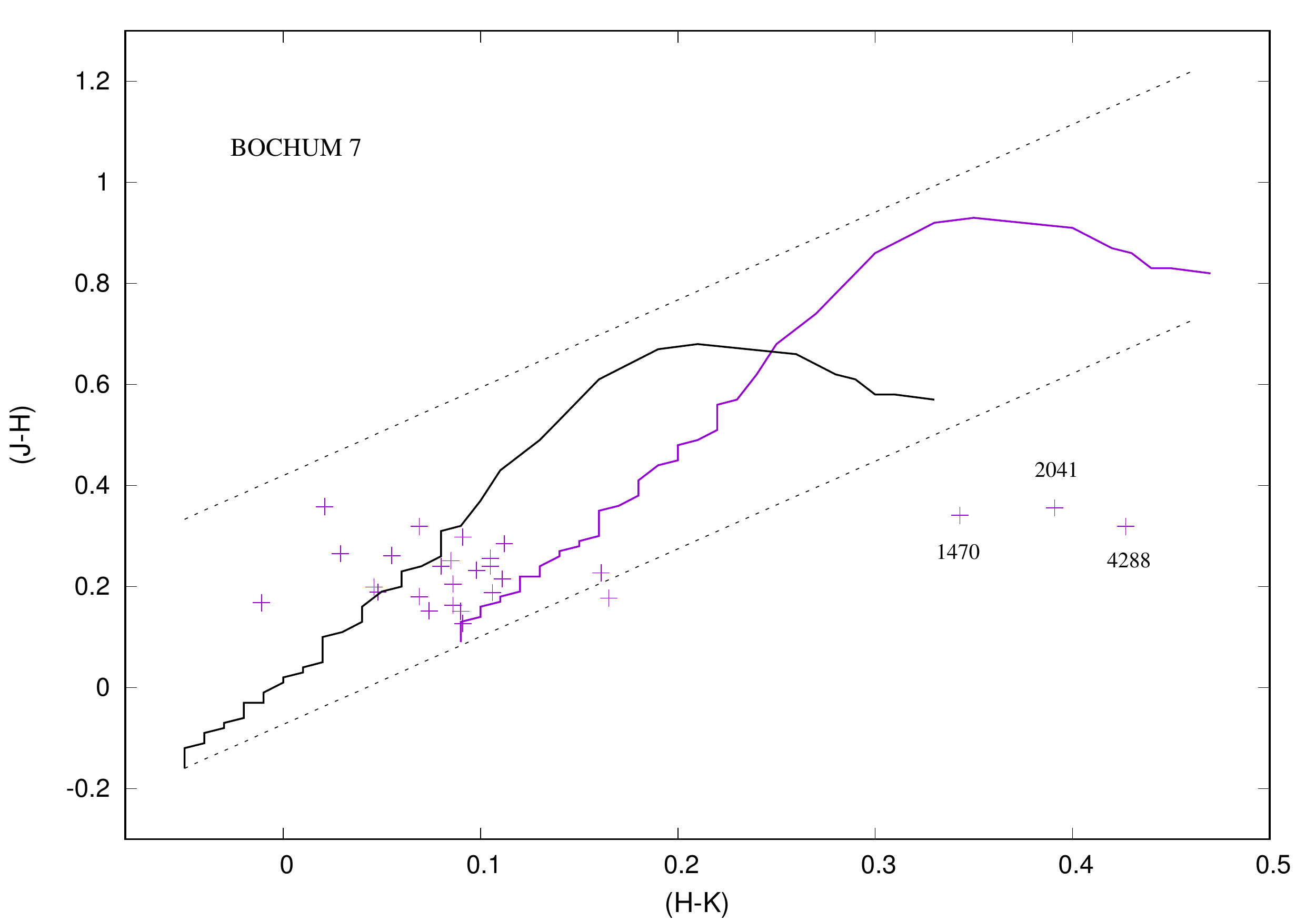}
\caption{Photometric diagrams for star members of Bo\,7. Dashed lines indicate the normal reddening path \citep{rie85}. The black and lilac solid curves are the unreddened and shifted curves, respectively. The lilac solid curve has been shifted according to the color excess MS \citep{koo83}. }
\label{c-cJHKBo7}
\end{figure}

The near-IR two-color diagram is a good tool for the selection of IR-excess stars. Figure \ref{c-cJHKBo7} is the $JHK$ color-color diagram obtained with 2MASS photometric data for member  stars of the Bo7 association; the black and lilac solid curves are the unreddened and shifted curves, respectively. The lilac solid curve has been shifted according to the color excess MS stars \citep{koo83}. The color excesses were obtained using the \citet{rie85} equations, knowing that the $E(B-V)$ is 0.77 mag for Bo 7.
The dashed lines indicate the normal reddening path \citep{rie85}. Next to the symbols, we show the stellar identification number from this paper (see Table \ref{PhotBo7}).

The error of $\simeq$ 0.02 mag in the $JHK$ star colors observed with 2MASS telescopes is likely the reason why some stars are so widely distributed around the MS curve. Bochum 7 has the stars \#1470, \#2041 and \#4288 that lie to the right of the normal reddening vector of an OB dwarf star. This apparent IR excess can be linked to the presence of current star formation in dusty disks or emission lines that enhance the observed K-band flux. Our optical spectra reveal that all these stars show emission lines (the star \#1470 is classified as B0\,IIIe, \#2041 as WN8 and \#4288 as B1-2Ve ) which can be responsible for the shift to the right in their (J-H) versus\ (H-K) diagram.

 The IRAS source 8426-4601 ({\it l, b}) = (265\fdg3,
-2\fdg2) is found in the vicinity of Bo\,7. In the survey of the CS\,(2-1) emission toward IRAS point sources in the galactic plane \citep{bro96} this IRAS source shows an emission line at radial velocity {\rm V} = 43.8 km s$^{-1}$. The color-color diagram obtained from the fluxes measured in the IR \citep{woo89} indicates that it could be an ultra-compact HII region. 
Molecular observations of different CO emission lines obtained with MOPRA\footnote{https://www.narrabri.atnf.csiro.au/mopra/} and APEX\footnote{http://www.eso.org/public/teles-instr/apex/} radio telescopes are currently being analyzed. The main objective of these observations is to characterize the molecular cloud physics associated with the source IRAS 8426-4601 and Bo 7 (paper in preparation).

\begin{table*}[t]
\caption{Spectrophotometric information of stars in Bo\,7, including distance modulus determination and measured radial velocities}
 \label{PhotBo7}
 \begin{scriptsize}
\centering
\begin{tabular}{crrrrrlccl}
 \hline \hline
   ID &  V & J & H & K & (B-V) & SpT & V$_o$ - M$_v$ & RV & Cross ref.  \\
         &             &       &         &        &              &        &    & (km s$^{-1}$) & CBN.........../2MASS.......... \\
 \hline
 343    &       14.8 &  13.5 &  13.3 &  13.1 &  0.59 &  b4 V    & 13.8 & -- & 84541.2-461113/J08454120-4611127 \\
 390 & 14.4 & 13.4 & 13.2 & 13.2 & 0.40 & b5 V & 13.8 & -- & 84538.9-455736/J08453890-4557364\\
 521    &       14.8 &  12.8 &  12.5 &  12.5 &  0.66 &  b5 V    & 13.6 & -- & 84535.3-460035/J08453526-4600351 \\
1031     &      14.8 &  13.3 &  13.0 &  12.9 &  0.57 &  B5 V & 13.5 & 49(16) (II) & 84518.8-454147/J08451888-4541494 \\
1052 &14.6 & 12.7 & 12.5 & 12.4 & 0.56 & b3 V & 13.7 & -- & 84519.0-460022/J08451895-4600221\\
1069 & 13.6 & 12.0 & 11.8  & 11.7 & 0.65 & B1 V & 13.8 & 88(33)(II) & 84518.2-455816/J08451822-4558166\\
1071     &      14.8 &  13.4 &  13.2  & 13.1 &  0.55 &  b5 V & 13.6 & -- & 84518.3-460042/J08451829-4600418 \\
1470 &  11.5 &  9.6 &           9.3 &           8.9     &        0.55 & B0 IIIe &  13.8 & var (Tab. \ref{tab47}) & 84506.4-460905/ALS\,1147 \\
1607     &      15.0 &  13.4 &  13.0 &  13.0 &  0.70 &  b3 V    & 13.6 & -- & 84502.2-460757/J08450226-4607572 \\
1939 &  11.6 &  10.3    &       10.1    &       10.1    &       0.47 &  B1 III & 13.6 & var (Tab. \ref{tab46}) & 84450.9-461012/ALS\,1146 \\
2041 & 10.8 & 8.6 & 8.3 & 7.9 & 0.51 & WN8 & 13.8 & -- & 84447.2-455856/ALS\,1145\\
2197 & 14.8 & 13.3 & 13.1 & 13.0 & 0.48 & b4V & 14.0 & -- & 84442.2-460900/J08444228-4609013 \\
2511    &       13.5 &  11.8 &  11.6 &  11.5 &  0.61    &        B2 III & 14.0 & 85(19) (II) & 84431.5-455314/J08443163-4553148\\
2720 &  11.3 &  9.8 &           9.6 &           9.4 &           0.61 &  O7.5 V &     13.6 & var (Tab. \ref{tab44}) & 84424.9-455334/ALS\,1144 \\
2781     &      13.7 &  11.8 &  11.6 &  11.4 &  0.65 &  B1-2/V-III &    13.9 & 55(32) (II) & 84423.1-460821/J08442325-4608192 \\
3211 & 14.9 & 13.3  & 13.1 & 13.0 & 0.58 & b6 V & 13.6 & -- & 84407.2-460155/J08440715-4601548 \\
3273 &  11.7 &  9.8 &           9.5 &   9.5 &           0.68 &  B0 III & 13.7 & var (Tab \ref{tab460746}) & 84404.5-455316/ALS\,1140 \\
3669 & 11.4 & 10.2 & 10.0 & 10.0 & 0.45 & O9-9.5 V & 13.6 & var (Tab \ref{tab460746}) & 84350.9-460348/ALS\,1137\\
3705 &  10.9 &          9.9 &           9.7 &           9.6 &           0.34 &       O6.5((f)) V &   14.1 & 71(1)$\star$ & 84349.6-460711/ALS\,1135\\
3767 & 15.0 & 13.2 & 13.0 & 13.0  & 0.56 & b6V & 13.7 & -- & 84347.5-460029/J08434760-4600562 \\
3829     &      12.9 &  11.7 &  11.5 &  11.4 &  0.43    &       B1 V    & 13.9 & 3(31) (II) & 84344.7-460656/J08434493-4606537\\
3982     &      12.0 &  10.7 &  10.5 &  10.5 &  0.45 &  O9.5 V & 13.6 & 56(17) (II)& 84338.5-460817/J08433863-4608157 \\
4009 & 14.2 & 12.7 & 12.5 & 12.4 & 0.47 & b4 V & 13.6 & -- & 84334.4-460553/J08433734-4605520\\
4234    &     12.6 &   11.2   &         11.0 &  10.9    &  0.56    &    B0 V & 13.8 &  55(27) (II) & 84327.4-460240/J08432763-4602382\\
4288 & 13.7 & 12.0 & 11.7 & 11.3 & 0.62 & B1-2 Ve & 13.6 & var (Tab \ref{tab460746}) & 84325.6-460820/J08432574-4608174\\
4331 & 10.8 & 9.8 & 9.7 & 9.6 & 0.46 & O7.5 V & 13.6 & var (Tab. \ref{tab31}) & 84324.0-460831/ALS\,1131\\
4476     &      14.5    &       13.1    &       12.9    &       12.8 &  0.38 &        b6 V & 13.9 & -- & 84317.8-460618/J08431784-4606180\\ 
\hline
\end{tabular}
\parbox{14cm}{{\scriptsize $\star$ \citet{cor03}}}\\
\end{scriptsize}
\end{table*}

\section{Summary}

Since the discovery of the Bo\,7 stellar group by \citet{mof75}, several studies have been carried out to identify its members  in the region using spectroscopic, photometric, and astrometric analysis techniques. This paper is the result of the first investigation carried out in the Bo\,7 region employing these astrophysics techniques simultaneously. With these techniques we confirmed that Bo\,7 is an OB association centered at  $\alpha$ = 8$^h$45$^m$, $\delta$ = $- 45^{\circ}$59' ($l = 265^{\circ}\!\!$.12, b = $-2^{\circ}\!\!$.0) at a distance of $\simeq$ 5.7 kpc and identified 27 stellar members. The components of the proper motion are $\mu_{\alpha}cos\delta$ = $-4.92 \pm 0.08$ mas yr$^{-1}$,  $\mu_{\delta}$ = $3.26 \pm 0.08$ mas yr$^{-1}$. 
Bochum\,7 is located on the edge of the HI supershell $GS263-02+45$ \citep{arn07} and perhaps the birth of the Bo\,7 association may be related to the evolution of this HI supershell.  Our results suggest that Bochum 7 has a heliocentric radial velocity of $\sim$ 35 km s$^{-1}$ (V$_{LSR}$ = 40 km s$^{-1}$ for d = 5.7 kpc, Fig. \ref{RotGalaxy}), and binary system has been confirmed (ALS\,1135, \citet{cor03}) as well as another four stars with variable radial velocities, thus making them  binary-system candidates.

\begin{acknowledgements}
This work was partially supported by Universidad Nacional de La Plata (UNLP) under projects 11/G144. This publication also made use of data from the Two Micron All
Sky Survey, which is a joint project of the University of Massachusetts and the Infrared Processing and Analysis Center/California Institute of Technology, funded by the National Aeronautics and Space Administration and the National Science Foundation (http://cdsportal.u-strasbg.fr/). This work has made use of data from the European Space Agency (ESA) mission {\it Gaia}(\url{https://www.cosmos.esa.int/gaia}), processed by
the {\it Gaia} Data Processing and Analysis Consortium (DPAC, (\url{https://www.cosmos.esa.int/web/gaia/dpac/consortium})). Funding for the DPAC has been provided by national institutions, in particular the institutions participating in the {\it Gaia} Multilateral Agreement.
We thank Jorge Panei for many useful comments which improved this paper. Finally, we wish to thank the anonymous referee for their suggestions and comments, which improved the original version of this work.
\end{acknowledgements}

\bibliographystyle{aa}  
\bibliography{bibliomar2018}

\begin{thebibliography}{39}
\expandafter\ifx\csname natexlab\endcsname\relax\def\natexlab#1{#1}\fi

\bibitem[{{Arnal} \& {Corti}(2007)}]{arn07}
{Arnal}, E.~M. \& {Corti}, M. 2007, \aap, 476, 255

\bibitem[{{Bagnuolo} {et~al.}(1999){Bagnuolo}, {Gies}, {Riddle}, \&
  {Penny}}]{bag99}
{Bagnuolo}, Jr., W.~G., {Gies}, D.~R., {Riddle}, R., \& {Penny}, L.~R. 1999,
  \apj, 527, 353

\bibitem[{{Bastian}(2011)}]{2011sca..conf...85B}
{Bastian}, N. 2011, in Stellar Clusters and Associations: A RIA Workshop on
  Gaia, 85--97

\bibitem[{{Bosch} {et~al.}(2009){Bosch}, {Terlevich}, \& {Terlevich}}]{bos09}
{Bosch}, G., {Terlevich}, E., \& {Terlevich}, R. 2009, \aj, 137, 3437

\bibitem[{{Brand} \& {Blitz}(1993)}]{bra93}
{Brand}, J. \& {Blitz}, L. 1993, \aap, 275, 67

\bibitem[{{Bronfman} {et~al.}(1996){Bronfman}, {Nyman}, \& {May}}]{bro96}
{Bronfman}, L., {Nyman}, L.-A., \& {May}, J. 1996, \aaps, 115, 81

\bibitem[{{Chen} {et~al.}(1997){Chen}, {Asiain}, {Figueras}, \&
  {Torra}}]{che97}
{Chen}, B., {Asiain}, R., {Figueras}, F., \& {Torra}, J. 1997, \aap, 318, 29

\bibitem[{{Corti} {et~al.}(2007){Corti}, {Bosch}, \& {Niemela}}]{cor07}
{Corti}, M., {Bosch}, G., \& {Niemela}, V. 2007, \aap, 467, 137

\bibitem[{Corti {et~al.}(2003)Corti, Niemela, \& Morrell}]{cor03}
Corti, M., Niemela, V., \& Morrell, N. 2003, A\&A, 405, 571

\bibitem[{{de Zeeuw} {et~al.}(1999){de Zeeuw}, {Hoogerwerf}, {de Bruijne},
  {Brown}, \& {Blaauw}}]{deze99}
{de Zeeuw}, P.~T., {Hoogerwerf}, R., {de Bruijne}, J.~H.~J., {Brown}, A.~G.~A.,
  \& {Blaauw}, A. 1999, \aj, 117, 354

\bibitem[{{Dias} {et~al.}(2002){Dias}, {Alessi}, {Moitinho}, \&
  {L{\'e}pine}}]{dia02}
{Dias}, W.~S., {Alessi}, B.~S., {Moitinho}, A., \& {L{\'e}pine}, J.~R.~D. 2002,
  \aap, 389, 871

\bibitem[{{Efremov} {et~al.}(1987){Efremov}, {Ivanov}, \& {Nikolov}}]{efr87}
{Efremov}, I.~N., {Ivanov}, G.~R., \& {Nikolov}, N.~S. 1987, \apss, 135, 119

\bibitem[{{Fari{\~n}a} {et~al.}(2009){Fari{\~n}a}, {Bosch}, {Morrell},
  {Barb{\'a}}, \& {Walborn}}]{far09}
{Fari{\~n}a}, C., {Bosch}, G.~L., {Morrell}, N.~I., {Barb{\'a}}, R.~H., \&
  {Walborn}, N.~R. 2009, \aj, 138, 510

\bibitem[{{Fern{\'a}ndez Laj{\'u}s} \& {Niemela}(2006)}]{fer06}
{Fern{\'a}ndez Laj{\'u}s}, E. \& {Niemela}, V.~S. 2006, \mnras, 367, 1709

\bibitem[{{Gaia Collaboration} {et~al.}(2016{\natexlab{a}}){Gaia
  Collaboration}, {Brown}, {Vallenari}, {Prusti}, {de Bruijne}, {Mignard},
  {Drimmel}, {Babusiaux}, {Bailer-Jones}, {Bastian}, \& et~al.}]{brow16}
{Gaia Collaboration}, {Brown}, A.~G.~A., {Vallenari}, A., {et~al.}
  2016{\natexlab{a}}, \aap, 595, A2

\bibitem[{{Gaia Collaboration} {et~al.}(2016{\natexlab{b}}){Gaia
  Collaboration}, {Prusti}, {de Bruijne}, {Brown}, {Vallenari}, {Babusiaux},
  {Bailer-Jones}, {Bastian}, {Biermann}, {Evans}, \& et~al.}]{pru16}
{Gaia Collaboration}, {Prusti}, T., {de Bruijne}, J.~H.~J., {et~al.}
  2016{\natexlab{b}}, \aap, 595, A1

\bibitem[{{Koornneef}(1983)}]{koo83}
{Koornneef}, J. 1983, \aap, 128, 84

\bibitem[{{Marigo} {et~al.}(2008){Marigo}, {Girardi}, {Bressan}, {Groenewegen},
  {Silva}, \& {Granato}}]{mari08}
{Marigo}, P., {Girardi}, L., {Bressan}, A., {et~al.} 2008, \aap, 482, 883

\bibitem[{{Michalska} {et~al.}(2013){Michalska}, {Niemczura}, {Pigulski},
  {Ste{\'s}licki}, \& {Williams}}]{mic13}
{Michalska}, G., {Niemczura}, E., {Pigulski}, A., {Ste{\'s}licki}, M., \&
  {Williams}, A. 2013, \mnras, 429, 1354

\bibitem[{Moffat \& Vogt(1975)}]{mof75}
Moffat, A. F.~J. \& Vogt, N. 1975, A\&AS, 20, 85

\bibitem[{Niemela(1982)}]{nie82}
Niemela, V.~S. 1982, In Wolf-Rayet Stars: Observations, Physics, Evolution,
  Proc. IAU Sump. 99 (eds. C. de Loore \& A. J. Willis), (Reidel - Dordrecht)

\bibitem[{{Orellana} {et~al.}(2010){Orellana}, {de Biasi}, {Bustos Fierro}, \&
  {Calder{\'o}n}}]{ore10}
{Orellana}, R.~B., {de Biasi}, M.~S., {Bustos Fierro}, I.~H., \&
  {Calder{\'o}n}, J.~H. 2010, \aap, 521, A39+

\bibitem[{{Orellana} {et~al.}(2015){Orellana}, {De Biasi}, {Pa{\'{\i}}z},
  {Bustos Fierro}, \& {Calder{\'o}n}}]{ore14}
{Orellana}, R.~B., {De Biasi}, M.~S., {Pa{\'{\i}}z}, L.~G., {Bustos Fierro},
  I.~H., \& {Calder{\'o}n}, J.~H. 2015, \na, 36, 70

\bibitem[{{Polcaro} {et~al.}(2016){Polcaro}, {Maryeva}, {Nesci}, {Calabresi},
  {Chieffi}, {Galleti}, {Gualandi}, {Haver}, {Mills}, {Osborn}, {Pasquali},
  {Rossi}, {Vasilyeva}, \& {Viotti}}]{pol16}
{Polcaro}, V.~F., {Maryeva}, O., {Nesci}, R., {et~al.} 2016, \aj, 151, 149

\bibitem[{{Reed}(2000)}]{ree00}
{Reed}, B.~C. 2000, \aj, 119, 1855

\bibitem[{{Rieke} \& {Lebofsky}(1985)}]{rie85}
{Rieke}, G.~H. \& {Lebofsky}, M.~J. 1985, \apj, 288, 618

\bibitem[{{Sanders}(1971)}]{san71}
{Sanders}, W.~L. 1971, \aap, 14, 226

\bibitem[{{Sano} {et~al.}(2017){Sano}, {Reynoso}, {Mitsuishi}, {Nakamura},
  {Furukawa}, {Mruganka}, {Fukuda}, {Yoshiike}, {Nishimura}, {Ohama}, {Torii},
  {Kuwahara}, {Okuda}, {Yamamoto}, {Tachihara}, \& {Fukui}}]{san17}
{Sano}, H., {Reynoso}, E.~M., {Mitsuishi}, I., {et~al.} 2017, Journal of High
  Energy Astrophysics, 15, 1

\bibitem[{Schmidt-Kaler(Th. 1982)}]{lando82}
Schmidt-Kaler. Th. 1982, In Landolt-Bornstein New Series, Group VI, Vol. 2b.
  (eds. K. Schaifers \& H. H. Voigt), (Springer-Verlag, Berlin)

\bibitem[{{Skrutskie} {et~al.}(2003){Skrutskie}, {Cutri}, {Stiening},
  {Weinberg}, {Schneider}, {Carpenter}, {Beichman}, {Capps}, {Chester},
  {Elias}, {Huchra}, {Liebert}, {Lonsdale}, {Monet}, {Price}, {Seitzer},
  {Jarrett}, {Kirkpatrick}, {Gizis}, {Howard}, {Evans}, {Fowler}, {Fullmer},
  {Hurt}, {Light}, {Kopan}, {Marsh}, {McCallon}, {Tam}, {van Dyk}, \&
  {Wheelock}}]{skr03}
{Skrutskie}, M.~F., {Cutri}, R.~M., {Stiening}, R., {et~al.} 2003, VizieR
  Online Data Catalog, 7233

\bibitem[{{Smith} \& {Conti}(2008)}]{smi08}
{Smith}, N. \& {Conti}, P.~S. 2008, \apj, 679, 1467

\bibitem[{Stenholm(1984)}]{lun84}
Stenholm, L. 1984, A\&AS, 58, 163

\bibitem[{Stephenson \& Sanduleak(1971)}]{ste71}
Stephenson, C. \& Sanduleak, N. 1971, Publ. Warner and Swasey Obs., 1, 1

\bibitem[{Sung {et~al.}(1999)Sung, Bessell, Park, \& Kang}]{sun99}
Sung, H., Bessell, M.~S., Park, B.~G., \& Kang, Y.~H. 1999, JKAS, 32, 109

\bibitem[{{van der Hucht}(2001)}]{vandh01}
{van der Hucht}, K.~A. 2001, New Astronomy Review, 45, 135

\bibitem[{{Vasilevskis} {et~al.}(1965){Vasilevskis}, {Sanders}, \& {van
  Altena}}]{vas65}
{Vasilevskis}, S., {Sanders}, W.~L., \& {van Altena}, W.~F. 1965, \aj, 70, 806

\bibitem[{{Walborn}(1972)}]{wal72}
{Walborn}, N.~R. 1972, \aj, 77, 312

\bibitem[{{Wood} \& {Churchwell}(1989)}]{woo89}
{Wood}, D.~O.~S. \& {Churchwell}, E. 1989, \apjs, 69, 831

\bibitem[{{Zacharias} {et~al.}(2017){Zacharias}, {Finch}, \& {Frouard}}]{zac17}
{Zacharias}, N., {Finch}, C., \& {Frouard}, J. 2017, \aj, 153, 166

\end{thebibliography}

\newpage

\begin{appendix}
\section{Individual stellar radial velocities}
For the determination of the radial velocities (Sect. \ref{stellarrv}) we analyzed digital spectra obtained by us for some stars present in the Bochum\,7 region. Details of the instrumental configuration employed to obtain the spectra are given in Sect. 2.2 of Paper\,I. Table \ref{ICU} of this paper lists some characteristics of the CASLEO \footnote{Visiting Astronomer, Complejo Astron\'omico El Leoncito operated under agreement between the Consejo Nacional de Investigaciones Cient\'{\i}ficas y T\'ecnicas de la Rep\'ublica Argentina and the National Universities of La Plata, C\'ordoba and San Juan.} runs. 

\begin{center}
\begin{table}
\caption{Instrumental configurations used.}
\label{ICU}
\begin{scriptsize}
\begin{tabular}{cllccc}
\hline
\hline
ID & Epoch(s)& Spectrog. & Rec.disp. & $\Delta\lambda$ & S/N\\
&            &          & (\AA~px$^{-1}$) & (\AA)&  \\
\hline
I & 1999 (Jan., Feb.) & B\&C & 2.30 &   3800-5000 & 120-200 \\
II & 2000 (Jan.) &  REOSC &  1.65 & 3700-5200 & 150-200 \\
I & 2001 (Jan.) & B\&C & 2.30 &   3800-5000 & 120-200 \\
II & 2001 (Feb.) &  REOSC &  1.65 & 3700-5200 & 150-200 \\
II & 2002 (Jan., Feb.) &  REOSC &  1.65 & 3700-5200 & 150-200 \\
\hline
\end{tabular}
\end{scriptsize}
\end{table}
\end{center} 

The radial velocity of each star was obtained measuring the absorption lines of H${\gamma}$, HeI~4471\AA\ , H${\beta}$ and HeI~4922\AA\ found in its spectra.
Probable members of Bo\,7 are shown in Table \ref{PhotBo7}. Here, we present the heliocentric radial velocity with the standard deviation of the radial velocity average and the instrumental configuration in two pairs of parentheses. There are several stars for which it was not possible to obtain a good estimate of radial velocity from spectra.
The ALS Catalog stars 1135 (CBN\,84349.6-460711), 1131 (CBN\,84324.0-460831), 1144 (CBN\,84424.9-455334), 1146 (CBN\,84450.9-461012) and 1147 (CBN\,84506.3-460906) show variations in radial velocity in their spectra. 

ALS\,1135 is a binary system presented by \citet{cor03}, \citet{fer06} and \citet{mic13}. The radial velocities measured in the other stars are presented from
Table \ref{tab31} to Table \ref{tab47}. They list Heliocentric Julian Day, instrumental configuration, heliocentric radial velocity (with the number of measured lines to average in parentheses) and in the last column, standard deviation of radial velocity average. The presence of systematic differences between the lines of spectra obtained from night to night could be due to a binary effect. ALS\,1131 is the star showing the greatest radial velocity dispersion with -45 km\,s$^{-1} \leq$ Vr $\leq$ 160 km\,s$^{-1}$. This star presents an additional characteristic typical of binary systems: the variation of the intensity ratio of HeI~4471 / HeII~4541, known as the Struve-Sahade effect \citep{bag99}. The intensity ratios obtained between the lines of HeI~4471 and HeII~4541 vary between 1.6 and 2.4. These data suggest this star to be a highly probable binary system, although more data with better resolution are needed. 
ALS\,1144 is another star showing a large radial velocity dispersion with -26 km\,s$^{-1} \leq$ Vr $\leq$ 100 km\,s$^{-1}$. ALS\,1146 shows radial velocity dispersion values somewhat smaller than the other  binary star candidates (80 km\,s$^{-1}$), so it could be a longer-period binary system. ALS\,1147 is a star with hydrogen Balmer lines in emission in its spectra and we found a large dispersion in its radial velocity values, flagging it as another probable binary system. 

The stars ALS\,1137 (CBN\,84350.9-460348), ALS\,1140 (CBN\,84404.4-455316), CBN\,84325.6-460820 and CBN\,84438.9-460746 also show variations in their radial velocity, and they are listed in Table \ref{tab460746}. For these four stars, more spectral data are required to confirm that the stellar radial velocity is indeed variable.

\begin{table}
\begin{center}
\leavevmode \caption{CBN84324.0-460831 ($ALS$\,1131): heliocentric radial velocities.}
\label{tab31} \vskip 0.2cm
\begin{small}
\begin{tabular}{cccc}
\hline \hline
$HJD$ & $IC$ & V$_r$ & $\sigma$ \\
(2450000+) &  & (km\,s$^{-1}$) & (km\,s$^{-1}$) \\
\hline
202.832 & I & 16(4) & $\pm$26 \\
1204.850 & I & 43(4) & $\pm$10 \\
1219.586 & I  & 52(4) & $\pm$15 \\
1555.641 & II & 18(4) & $\pm$16 \\
1556.733 & II & 47(4) & $\pm$29 \\
1557.773 & II & -21(4) & $\pm$04 \\
1558.649 & II & 71(4) & $\pm$21 \\
1559.641 & II & 51(4) & $\pm$11 \\
1560.671 & II & 49(4) & $\pm$08 \\
1561.636 & II & 48(4) & $\pm$05 \\
1565.707 & II & 121(4) & $\pm$10 \\
1572.809 & II & 20(4) & $\pm$08  \\
1573.750 & II & 51(4) & $\pm$10 \\
1574.782 & II & 15(4) & $\pm$23 \\
1924.712 & I & 160(4) & $\pm$44 \\
1927.845 & I & -23(4) & $\pm$34 \\
1945.806 & II & 25(4) & $\pm$33 \\
1946.575 & II & 106(4) & $\pm$16 \\
2293.774 & II & -45(4) & $\pm$16 \\
2296.861 & II & 28(4) & $\pm$39 \\
2298.726 & II & 38(4) & $\pm$19 \\
2299.725 & II & 130(4) & $\pm$25 \\
2300.699 & II & 30(4) & $\pm$24 \\
2301.674 & II & 60(4) & $\pm$34 \\
2302.671 & II & 33(4) & $\pm$28 \\
2322.733 & II & -16(4) & $\pm$07  \\
2323.614 & II & 87(4) & $\pm$04  \\
2324.538 & II & 74(4) & $\pm$09  \\
\hline
\end{tabular}
\end{small}
\end{center}
\end{table}

\begin{table}
\begin{center}
\leavevmode \caption{CBN84424.9-455334 ($ALS$\,1144): heliocentric radial velocities.}
\label{tab44} \vskip 0.2cm
\begin{small}
\begin{tabular}{cccc}
\hline \hline
$HJD$ & $IC$ & V$_r$ & $\sigma$ \\
(2450000+) &  & (km\,s$^{-1}$) & (km\,s$^{-1}$) \\
\hline
1203.793 & I & 48(4) & $\pm$15 \\
1219.604 & I & 26(4) & $\pm$04 \\
1221.834 & I & 82(4) & $\pm$15 \\
1555.834 & II & 60(4) & $\pm$14 \\
1556.720 & II & 29(4) & $\pm$13 \\
1557.818 & II & 19(4) & $\pm$16 \\
1558.671 & II & 46(4) & $\pm$19 \\
1559.666 & II & 66(4) & $\pm$15 \\
1560.711 & II & 64(4) & $\pm$10 \\
1561.647 & II & 18(4) & $\pm$09  \\
1565.658 & II & 37(4) & $\pm$05  \\
1566.781 & II & 00(4) & $\pm$18 \\
1573.761 & II & 56(4) & $\pm$03 \\
1574.767 & II & 26(3) & $\pm$07 \\
1575.853 & II & -26(4) & $\pm$22 \\
1924.867 & I & 87(4) & $\pm$14 \\
1925.807 & I & 55(4) & $\pm$19 \\
1927.863 & I & 42(4) & $\pm$15 \\
1945.824 & II & 62(4) & $\pm$15 \\
1946.560 & II & 39(4) & $\pm$12 \\
2293.672 & II & 100(4) & $\pm$10 \\
2296.687 & II & 23(4) & $\pm$16 \\
2323.594 & II & 60(4) & $\pm$04 \\
2324.556 & II & 77(4) & $\pm$13 \\
\hline
\end{tabular}
\end{small}
\end{center}
\end{table}

\begin{table}
\begin{center}
\leavevmode \caption {CBN84450.9-461012 ($ALS$\,1146): heliocentric radial velocities.}
\label{tab46} \vskip 0.2cm
\begin{small}
\begin{tabular}{cccc}
\hline\hline
$HJD$ & $IC$ & V$_r$ & $\sigma$ \\
(2450000+) &  & (km\,s$^{-1}$) & (km\,s$^{-1}$) \\
\hline
1203.729 & I & 16(4) & $\pm$11 \\
1205.772 & I & -04(4) & $\pm$05 \\
 1220.841 & I & 47(4) & $\pm$32 \\
 1555.755 & II & 36(4) & $\pm$11 \\
 1556.745 & II & 48(4) & $\pm$12 \\
 1557.761 & II & 49(4) & $\pm$19 \\
 1558.625 & II & 27(4) & $\pm$24 \\
 1559.628 & II & 37(4) & $\pm$27 \\
 1560.658 & II & 38(4) & $\pm$15 \\
 1561.624 & II & 39(4) & $\pm$38 \\
 1565.689 & II & 26(4) & $\pm$15 \\
 1570.853 & II & 23(4) & $\pm$24 \\
 1572.800 & II & 12(4) & $\pm$19 \\
 1574.687 & II & 26(4) & $\pm$20 \\
 1924.785 & I & 41(4) & $\pm$23 \\
 1926.756 & I & 37(3) & $\pm$23 \\
 1946.591 & II & 76(4) & $\pm$25 \\
\hline
\end{tabular}
\end{small}
\end{center}
\end{table}

\begin{table}
\begin{center}
\leavevmode \caption{CBN84506.3-460906 ($ALS$\,1147): heliocentric radial velocities.}
\label{tab47} \vskip 0.2cm
\begin{small}
\begin{tabular}{cccc}
\hline \hline
$HJD$ & $IC$ & V$_r$ & $\sigma$ \\
(2450000+) &  & (km\,s$^{-1}$) & (km\,s$^{-1}$) \\
\hline
 1203.752 & I & -32(3) & $\pm$30 \\
 1205.788 & I & 03(3) & $\pm$19  \\
 1219.654 & I & -04(3) & $\pm$07 \\
 1220.857 & I & 26(3) & $\pm$10 \\
 1221.848 & I & -02(4) & $\pm$21 \\
 1555.770 & II & 12(3) & $\pm$25 \\
 1556.757 & II & 15(3) & $\pm$24 \\
 1557.750 & II & 25(3) & $\pm$03 \\
 1558.636 & II & 57(3) & $\pm$34 \\
 1559.617 & II & 11(3) & $\pm$07 \\
 1560.623 & II & 11(3) & $\pm$29 \\
 1561.613 & II & 19(3) & $\pm$17\\
 1565.670 & II & -21(3) & $\pm$33 \\
 1566.792 & II & 00(3) & $\pm$20 \\
 1572.790 & II & 15(3) & $\pm$21 \\
 1575.865 & II & 02(3) & $\pm$48 \\
 1924.760 & I & -18(3) & $\pm$11 \\
 1926.735 & I & 14(3) & $\pm$28 \\
 1946.606 & II & 14(4) & $\pm$37 \\
\hline
\end{tabular}
\end{small}
\end{center}
\end{table}

\begin{table}
\begin{center}
\leavevmode \caption{Heliocentric radial velocities.}
\label{tab460746} \vskip 0.2cm
\begin{small}
\begin{tabular}{lrcrc}
\hline \hline
ID & $HJD$ & $IC$ & V$_r$ & $\sigma$ \\
     & (2450000+) &  & (km\,s$^{-1}$) & (km\,s$^{-1}$) \\
\hline
CBN84438.9-460746 & 1558.779 & II & 49(4) & $\pm$18 \\
                                & 1946.654 & II & 73(4) & $\pm$22 \\
\hline
CBN84325.6-460820 & 925.730 & I & -13(4) & $\pm$36 \\
                                & 2293.774 & II & -27(4) & $\pm$32 \\
\hline
CBN84350.9-460348  & 1202.814& I & 57(4) & $\pm$16 \\
                & 1204.835 & I & 37(4) & $\pm$24 \\
                & 1558.660 & II & 72(4) & $\pm$09 \\
\hline
CBN84404.5-455316 & 203.773 & I & 40(4) & $\pm$10 \\
                                & 1219.826 & I & 53(4) & $\pm$21 \\
                                & 1558.682 & II & 74(4) & $\pm$17 \\
\hline
\end{tabular}
\end{small}
\end{center}
\end{table}
\end{appendix}

\IfFileExists{\jobname.bbl}{}
\typeout{}
\typeout{****************************************************}
\typeout{****************************************************}
\typeout{** the bibliography and then re-run LaTeX}
\typeout{** twice to fix the references!}
\typeout{****************************************************}
\typeout{****************************************************}
\typeout{}

\end{document}